\documentclass[pra,twocolumn,showpacs,floatfix]{revtex4}

\usepackage{amsmath}
\usepackage{times}
\usepackage{txfonts}
\usepackage{graphicx}
\usepackage{color}
\usepackage[colorlinks,linkcolor=blue,citecolor=blue,urlcolor=blue]{hyperref}
\begin{document}

\title{Interactions and Collisions of Discrete Breathers in \\
Two-Species Bose-Einstein Condensates in Optical Lattices} 

\author{Russell \surname{Campbell}}
\author{Gian-Luca \surname{Oppo}} 
\affiliation{Institute of Complex Systems, SUPA and Department of Physics, 
University of Strathclyde, 107 Rottenrow, Glasgow G4 0NG, Scotland, UK.}
\email{russell.campbell@strath.ac.uk}  

\author{Mateusz \surname{Borkowski}} 
\affiliation{Instytut Fizyki, Uniwersytet Miko\l{}aja Kopernika, 
  ul. Grudzi\c{a}dzka 5/7, 87--100 Toru\'n, Poland.}

\date{\today}

\begin{abstract}
The dynamics of static and travelling breathers in two-species Bose-Einstein 
condensates in a one-dimensional optical lattice is modelled within the 
tight-binding approximation. Two coupled discrete nonlinear Schr\"odinger 
equations describe the interaction of the condensates in two cases of relevance:
a mixture of two ytterbium isotopes and a mixture of $^{87}$Rb and $^{41}$K. 
Depending on their intial separation, interaction between static breathers of 
different species can lead to the formation of symbiotic structures and 
transform one of the breathers from a static into a travelling one.  
Collisions between travelling and static discrete breathers composed of 
different species are separated in four distinct regimes ranging from totally 
elastic when the interspecies interaction is highly attractive to mutual 
destruction when the interaction is sufficiently large and repulsive. We 
provide an explanation of the collision features in terms of the interspecies 
coupling and the negative effective mass of the discrete breathers.
\end{abstract}


\maketitle


\section{Introduction}

Bose-Einstein condensates have become a formidable tool for studying basic 
fundamentals of atomic physics \cite{dalfovo99,pitaevskii04}. When confined 
to an optical lattice they serve as an interesting analogue to a solid-state
system \cite{morsch06, bloch08}, providing means to study solid-state phenomena 
at an unprecedented level of parameter control. These include quantum phase 
transitions from the superfluid to Mott-insulator regimes \cite{greiner02}, 
transport phenomena \cite{chiofalo01}, Anderson localization \cite{roati08}, 
low-dimensional systems \cite{pethick13}, discrete breathers \cite{franzosi11} 
and solitons \cite{strecker02}. The latter can exist both with or without a 
periodic potential, although a lattice environment makes it possible for solitons 
and discrete breathers to exist even when the interactions in the BEC are 
repulsive (``gap solitons'') \cite{trombettoni01, abdullaev01, cruz09}. Such 
states have been observed experimentally in \cite{eiermann03}. Several other 
methods have been proposed in the literature, including the use of an additional 
harmonic potential \cite{matuszewski06} and boundary dissipations \cite{livi06}, 
which could also be used for non-demolition probing of these states 
\cite{franzosi07}.

The introduction of a second atomic species, i.e. the creation of a binary 
mixture of Bose-Einstein Condensates leads to even richer physics. In a 
harmonic trap the two species may be immiscible due to the interspecies
interaction, leading to phase separation \cite{hall98,ohberg98}. In an
effectively one-dimensional environment, the repulsive interaction
between atoms of different species (`interspecies' interactions)
during the formation of the condensate can leave the mixture far from
its ground state \cite{papp08,ronen08}. The interspecies interactions 
heavily influence the transport properties of a condensate in a lattice 
\cite{ruostekoski07, hooley07}. Repulsive interactions enable
the formation and extend the stability region of the so-called
\emph{symbiotic} gap solitons \cite{gubeskys06, adhikari08, malomed12},
i.e. two-species solitons localized together in the same spot of the
lattice. Somewhat counter-intuitively, an attractive interspecies
interaction may split the two overlapping solitons \cite{matuszewski07}.

So far most of the work has concentrated on the existence of two-species 
solitons and their stability \cite{abdullaev08, adhikari08, ali09, cruz07, 
gubeskys06, malomed12, matuszewski07, shi08}. In this paper we take a 
complementary approach -- we start with well-defined single-species discrete 
solitons (two travelling breathers or a self-trapped state and a travelling 
breather, the first made of one species and the other of the second species) 
far apart in the lattice and simulate their collisions. For a single species 
BEC confined to a one-dimensional lattice collisions of travelling breathers 
have been analyzed in detail in \cite{dabrowska04}. Such an approach has also 
been used in the case of binary BECs in harmonic traps \cite{busch01, zhang09}.
We perform our simulations using experimentally reachable conditions, with 
specific reference to two feasible experiments of two-species BECs in optical 
lattices.

The first of these experiments of interest has been performed in Kyoto where 
BECs of Yb atoms have been obtained separately with isotopes $^{174}$Yb 
\cite{takasu03} and $^{170}$Yb \cite{fukuhara07a}, as well as a mixture of 
isotopes $^{174}$Yb and $^{176}$Yb \cite{fukuhara08} (see also \cite{kasamatsu08}). 

In the mixture case, however, the latter
component quickly collapses due to its negative intraspecies
scattering length. The intra- and interspecies scattering lengths of
ytterbium isotopes have been measured using data from one-
\cite{enomoto07} and two-color photoassociation spectroscopy
\cite{kitagawa08} and are now well established. The rich isotope structure
of ytterbium enables mass tuning of the scattering length. It has been
shown that, unlike in alkali-metal species, optical Feshbach resonances 
can be used to effectively change the intraspecies scattering length 
\cite{ciurylo05,enomoto08,borkowski09}, thus raising hope for
optical control of interactions between different isotopes. Three
bosonic isotopes of ytterbium, namely $^{168,170,174}$Yb have positive
intraspecies scattering lengths of the order of a few nanometers
leading to similar stable condensates.  Since the isotope shifts are
small compared to the detuning of the far off resonant trap (FORT),
the potential seen by different isotopes is basically identical.

Even though these three isotopes are similar in terms of the
single-species scattering length, the interspecies interactions of
different pairs of isotopes differ dramatically. The interaction
between a $^{170}$Yb and $^{174}$Yb atoms is described by a large
negative scattering length of -27.3~nm, while for $^{168}$Yb and
$^{170}$Yb it is positive and equal to 6.2~nm. Halfway between these
two is the case of $^{168}$Yb and $^{174}$Yb characterised by a
negligible scattering length of 0.13(18)~nm, where the two condensate
species should essentially ignore each other.

Recently, an interesting mixture of heteronuclear BECs has been
obtained in an experiment of Thalhammer et al. \cite{thalhammer08},
where $^{41}$K and $^{87}$Rb atoms are condensed together in an
optical lattice. A remarkable property of this mixture is that the
interspecies scattering length $a_{1,2}$ describing the effective
interaction of colliding potassium and rubidium atoms can be tuned
over a wide (both positive and negative) range using a magnetic
Feshbach resonance, while the single-species scattering length remains
positive for either species. 

We present an analysis of the interaction and collisional behavior of discrete
breathers in two-species BEC in optical lattices in the tight-binding
approximation which has been successfully used to describe single-species 
experiments \cite{trombettoni01,abdullaev01,franzosi11,neff14}.  
In Section II the derivation of the model and an estimate of the
parameters are provided. Interaction of stationary breathers in close 
proximity to each other is described in section III.  Collision of 
traveling breathers and trapped
states are described in Section IV as a function of the interspecies
coupling parameter. Section V provides an explanation of the different 
kind of collisions observed in the numerical simulations where inelastic 
behavior is found in the mutually-repulsive case and elastic in the
mutually-attractive case. 



\section{The tight-binding approximation applied to a two-species Bose gas}

We analyze the behavior of the two-species Bose gas with the use of
the tight-binding approximation, following the treatment described
in \cite{ruostekoski07}. The time-dependent Gross-Pitaevskii equations 
describing the dynamics of the two species' order parameters $\Psi_i$ 
(where $i=1,2$) read
\begin{equation}
  i \hbar \dot \Psi_i(\vec r) = \left(-\frac{\hbar^2}{2m_i}\nabla^2 + U(\vec r) 
  + \sum_{j=1,2} g_{i,j} |\Psi_j(\vec r)|^2 \right) \Psi_i (\vec r) \, ,
  \label{GPEs}
\end{equation}
where the coefficients $g_{i,j}$ describe the effective mean-field 
intra- and interspecies interactions and are given by
%
%
\begin{equation}
  g_{i,j} = \frac{4\pi \hbar^2 a_{i,j}}{2\mu_{i,j}} \, ,
  \label{GCoeff}
\end{equation}
where $\mu_{i,j} = (m_i^{-1} + m_j^{-1})^{-1}$ is the reduced mass of the 
atomic pair and $a_{i,j}$ is the scattering length relevant in the scattering 
properties of the species' atoms.

The external potential confining the BECs is due to two overlapping
and counter-propagating laser beams that create a standing wave in
the axial direction and, as a result, a periodic potential of depth
$V_{0,i}$. The Gaussian profile of the two laser beams gives rise to
an approximately harmonic off-axis confinement described by the
frequencies $\omega_{r,i}$. Thus the external potential reads
\begin{equation}
  V_i = V_{0,i} \sin^2(kx) + \frac{1}{2}m_i \omega_{r,i}^2 (y^2+z^2) \, .
  \label{ExtPot}
\end{equation}
Note that real optical lattices also have a shallow harmonic
potential superimposed in the axial direction, which we here assume to
have negligible effects. We also introduce the lattice strengths 
$s_i = V_{0,i} / E_{R, i}$, where $E_{r,i} = {\hbar^2k^2}/{2m_i}$ is the 
recoil energy calculated for the lattice wavelength. The axial on-site 
frequency of the lattice is then $\omega_i = \sqrt{s_i} \hbar k^2 / m_i$.

Both order parameters obey the normalisation condition 
$\int {\rm d}^3 \vec r |\Psi_i|^2 = N_i$, where $N_i$ denotes the total 
number of atoms of the $i$-th species in the lattice. 

If the lattice is strong enough, i.e. the trap depth is sufficiently
large, the condensate is well localised around potential minima. For
each of the mixture's order parameters we use the following
\emph{ansatz} \cite{trombettoni01,ruostekoski07,abdullaev08}:
\begin{equation}
  \Psi_i = \sum_n \psi_{i,n}(t) \phi_i(\vec r - \vec r_{i,n}) \, ,
  \label{ansatz}
\end{equation}
where $\phi_i$ is an on-site wavefunction and $\vec r_{i,n}$
is the location of the $n$-th lattice site seen by the $i$-th
species. When the atomic interactions are weak, the on-site ground
state wavefunction can be replaced by the ground state harmonic oscillator
wavefunction in the off-axis direction and a Wannier function \cite{kohn59} 
of the lowest band in the axial direction to account for tunneling. 
Consequently, $|\psi_{i,n}(t)|^2$ may be interpreted as the number of $i$-th 
species atoms in the $n$-th lattice site as a function of time and $\sum_n
|\psi_{i,n}|^2 = N_i $. 

Substituting this \emph{ansatz} into the Gross-Pitaevskii equations and
dropping all terms mixing different lattice sites except for the ones
that describe tunneling (see \cite{trombettoni01, ruostekoski07}) one obtains
\begin{eqnarray}
  i\hbar \dot \psi_{i,n} & = & -J_i \left(\psi_{i,n-1}+\psi_{i,n+1} \right)  
\nonumber \\
  & &  + \left(\lambda_{i,i}|\psi_{i,n}|^2 + \lambda_{1,2}|\psi_{3-i,n}|^2 + 
\epsilon_{i,n} \right) \psi_{i,n}
\end{eqnarray}
where 
\begin{equation}
  J_i = \int {\rm d}^3 \vec r \phi_i(\vec r - \vec r_{i,n}) \left (
\frac{-\hbar^2}{2m_i}\nabla^2 + V_{\rm i} \right ) \phi_i(\vec r - \vec r_{i,n+1})
\end{equation}
is the hopping integral describing the tunneling of the
$i$-th species and is proportional to the intersite tunneling
rate $\gamma_i = J_i/\hbar$, while $\epsilon_{i,n} = \int {\rm d}^3 \vec r
\phi_i(\vec r-\vec r_{i,n})\left( \frac{-\hbar^2}{2m_i}\nabla^2 + V_{i}\right) 
\phi_i(\vec r-\vec r_{i,n})$ is the on-site chemical
potential. The self- and mutual interaction is described by the
parameters $\lambda_{i,i} = g_{i,i} \int{\rm d}^3 \vec r |\phi_i(\vec
r)|^4$ and $\lambda_{1,2} = g_{1,2} \int{\rm d}^3 \vec r |\phi_1(\vec
r)|^2|\phi_2(\vec r)|^2$, respectively.

\subsection{Normalization}
In order to move towards a more standard and computationally efficient form of 
two coupled discrete nonlinear Schr\"odinger equation (DNLSE), we introduce
\begin{eqnarray}
    z_{i,n} &=& \sqrt{\frac{1}{N_i}} \psi^*_{i,n} \, \exp \left( 
    -i \frac{\epsilon_{i} \tau} {\hbar \gamma_1} \right) \\
    \tau &=& \gamma_1 t 
\end{eqnarray}
to obtain:
\begin{eqnarray}
  \label{CDNLS1}
  i \frac{d}{d\tau} z_{1,n} &=& \Lambda_{1,1} |z_{1,n}|^2 z_{1,n} + \Lambda_{1,2} 
  \frac{N_2}{N_1} |z_{2,n}|^2 z_{1,n} \nonumber \\
  &-& z_{1,n-1}-z_{1,n+1} \\
  \label{CDNLS2}
  i \frac{d}{d\tau} z_{2,n} &=& \Lambda_{2,2} |z_{2,n}|^2 z_{2,n} + \Lambda_{1,2} 
  |z_{1,n}|^2 z_{2,n} \nonumber \\
  &-& \frac{\gamma_2}{\gamma_1} \left( z_{2,n-1}+z_{2,n+1} \right ) \, .
\end{eqnarray}
In Eqs. (\ref{CDNLS1}-\ref{CDNLS2}) we have defined the following parameters:
\begin{eqnarray}
  \Lambda_{i,i} = \frac{\lambda_{i,i} N_k}{\hbar \gamma_1 } 
  \;\;\;\;\;\;\;\;\;\;\;\;
  \Lambda_{1,2} = \frac{\lambda_{1,2} N_1}{\hbar \gamma_1 } \, .
\end{eqnarray}
Thus the atomic distribution of each species over the entire lattice is 
normalized to unity:
\begin{equation}
  \sum_n |z_{i,n}|^2 = 1 \, .
\end{equation}
To ensure that the energy and density in the system are conserved, we use a symplectic 
fourth-order integrator of the Yoshida type \cite{yoshida90,iubini13}. 
The energy and density are both conserved up to 9 decimal places at each integration 
time step.


\subsection{Estimate of the calculation parameters} 

  \begin{table}
    {\color{black}
    \caption{Values of parameters used in the simulations\label{parametertable}}
    \begin{tabular}{c c c c c c c c}
      \hline
      \hline
      Pair &$s_1$ & $s_2$ & $\gamma_2/\gamma_1$ & $\Lambda_{1,1}$ & $\Lambda_{2,2}$ & $\Lambda_{1,2}$\\
      \hline
      $^{168}$Yb + $^{170}$Yb & 3.19 & 3.27 & 0.96 & 5.28 & 1.368 & 2.51\\
      $^{170}$Yb + $^{174}$Yb & 3.27 & 3.43 & 0.91 & 1.368 & 2.486 & --12.24\\
      $^{87}$Rb + $^{41}$K    & 3.03 & 7    & 6.97 & 12.31 & 5.89 & (free)\\
      \hline
    \end{tabular}}
  \end{table}

In Section III(A) we present results that model different Bose gas
mixtures, notably mixtures of ytterbium isotopes and that of
$^{41}$K+$^{87}$Rb. At present, the only ytterbium isotope mixture BEC
obtained so far is that of $^{174}$Yb+$^{176}$Yb where, however, the
$^{176}$Yb part instantly collapses because of its negative scattering
length \cite{fukuhara08}. Thus we focus on ytterbium mixtures of
isotopes whose scattering length is positive, namely $^{168}$Yb,
$^{170}$Yb and $^{174}$Yb. The latter two have already reached BEC
separately \cite{takasu03,fukuhara07a} while the major technical
difficulty in reaching a $^{168}$Yb BEC is its low natural abundance
of 0.13\%.

To calculate the self nonlinear parameter, $\lambda_{i,i}$, it is sufficient 
to approximate the on-site wavefunction $\phi_{i,n}(\vec r)$ with a harmonic 
oscillator ground state, which is basically a Gaussian, to yield
\begin{equation}
  \lambda_{i,i}= a_{i,i}\sqrt{\frac{m \omega_r^2 \omega}{2 \pi \hbar}}
\end{equation}
In the case of ytterbium mixtures the mutual interaction parameter $\lambda_{1,2}$ 
can also be estimated from the above formula as the masses of the two isotopes 
are very similar and the two wavefunctions are well overlapped. 

In the case of the $^{41}$K+$^{87}$Rb mixture the two wavefunctions are differently 
shaped (one is narrower than the other due to the difference in masses). The difference 
in the masses of the two species has a further effect - the two clouds are separated 
due to gravity but an additional laser can be used to force overlap of the two species
\cite{inguscioprivate}. The scattering length of $^{87}$Rb is 5.25~nm \cite{marte02} 
while for $^{41}$K it is 3.1~nm \cite{modugno02,wang00}. The interspecies scattering 
length can be changed by the use of a convenient magnetic Feshbach resonance 
\cite{catani08} so that $\lambda_{1,2}$ can be considered as a free parameter.

To estimate the tunneling rates $\gamma_i$, it is not sufficient to approximate the 
on-site wavefunctions with Gaussians and one has to use the Wannier wavefunctions. 
This is because the tunneling rate is mostly determined by the wings of the on-site 
wavefunction which have an oscillatory-exponential rather than a Gaussian tail.  
In this case the tunneling rate can be approximated as \cite{bloch08}
\begin{equation}
    J_i = \hbar \gamma_i = \frac{4}{\sqrt{\pi}}E_{r,i}s_i^{3/4}\exp(-2\sqrt{s_i}) \, .
\end{equation}
The parameters for the ytterbium isotopes considered here are as follows. The lattice 
laser wavelength is $532$~nm; the lattice frequencies for both species are 
$\omega_r = 2\pi \times 100$~Hz and $\omega = 2\pi \times 15$~kHz which are close 
to the experimental realizations. For the sake of simplicity, we consider 
$N_1=N_2=O(10^3)$ as it is difficult to find initial conditions that would lead 
to a \emph{clear} traveling breather at larger densities. This is a general property 
of gap solitons \cite{matuszewski06}. The results apply, however, to larger values of 
$N_i$ up to $10^5$.

In the case of the $^{87}$Rb and $^{41}$K mixture we consider a lattice wavelength of 
1064~nm and we take the (tunable) lattice depth to 
be $V_e = 7 E_r$ for rubidium so that the same parameter for potassium is around 3. 
This is done to ensure that we can still use the tight-binding approximation (i.e. 
tunneling rates to further sites are at least an order of magnitude smaller than 
$\gamma_i$, see table I in \cite{bloch08}). On the other side the tunneling rate needs 
to be large enough for traveling breathers to exist. 

The calculation parameters discussed above are shown in table \ref{parametertable}.


\section{Breather interaction}

\begin{figure*}[t]
\includegraphics[angle=0.0,clip,width=1.8\columnwidth]{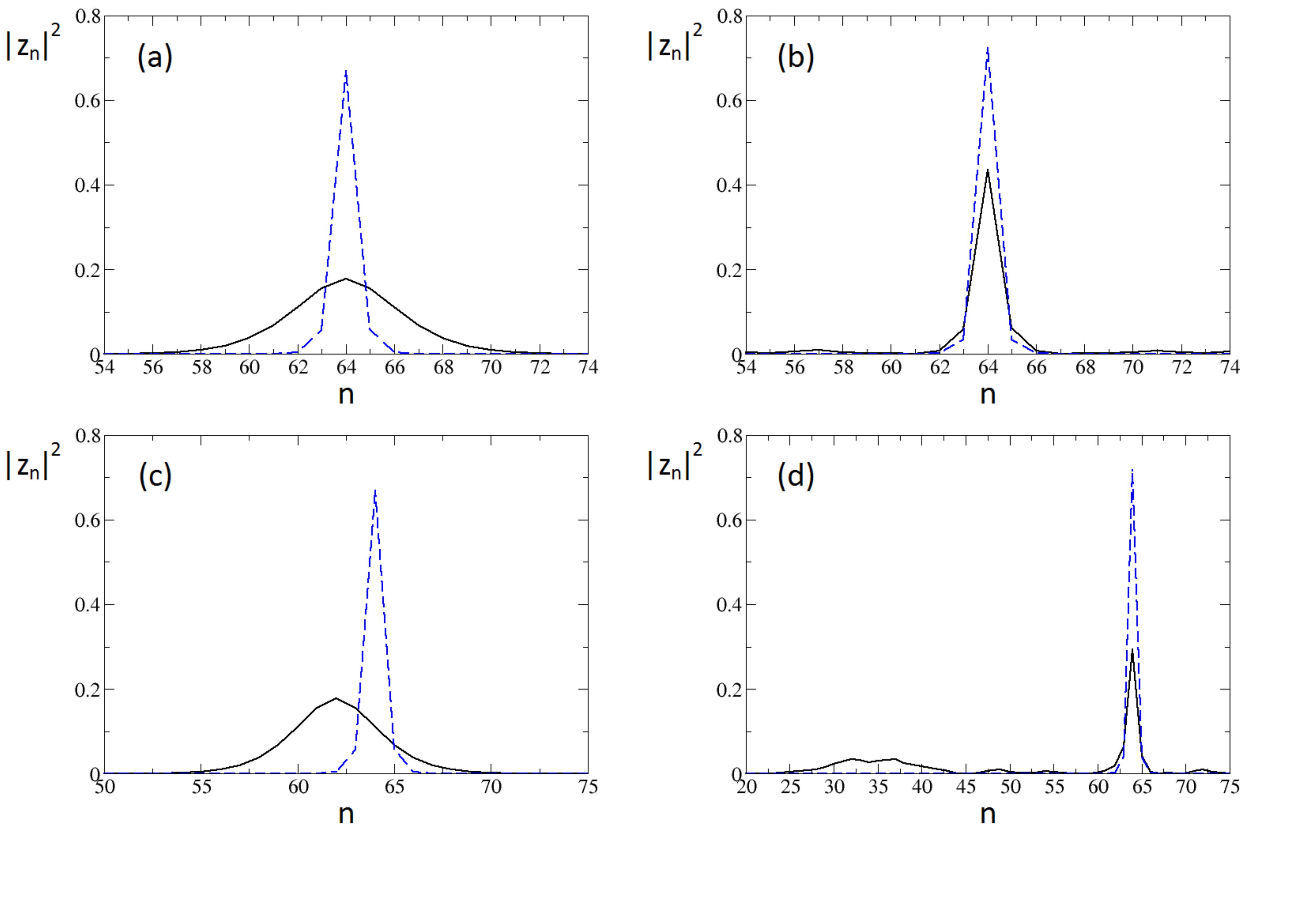}
\caption{\color{black}(Color online) Density profiles of $^{168}$Yb (black 
solid line) + $^{170}$Yb (blue dashed line) mixture. (a) Initial density 
profile of $^{168}$Yb + $^{170}$Yb breathers with $D=0$. (b) Density 
profile of $^{168}$Yb + $^{170}$Yb breathers at $\tau=1000$ after $\Lambda_{1,2}$
is switched on for $D=0$.  Note that the density profiles have changed 
when forming the symbiotic breather. (c) Initial density profile of 
$^{168}$Yb + $^{170}$Yb breathers with $D=2$. The wavefunctions of the two 
species still overlap significantly.  (d) Density profile of 
$^{168}$Yb + $^{170}$Yb breathers at $\tau=1000$ 
after $\Lambda_{1,2}$ is switched on for $D=2$.  Note that the $^{168}$Yb 
is smaller and some of the background has become localised to the left of 
the main breather.\label{figuredensprof}}
\end{figure*}

\begin{figure}[t]
\includegraphics[angle=0.0,clip,width=0.9\columnwidth]{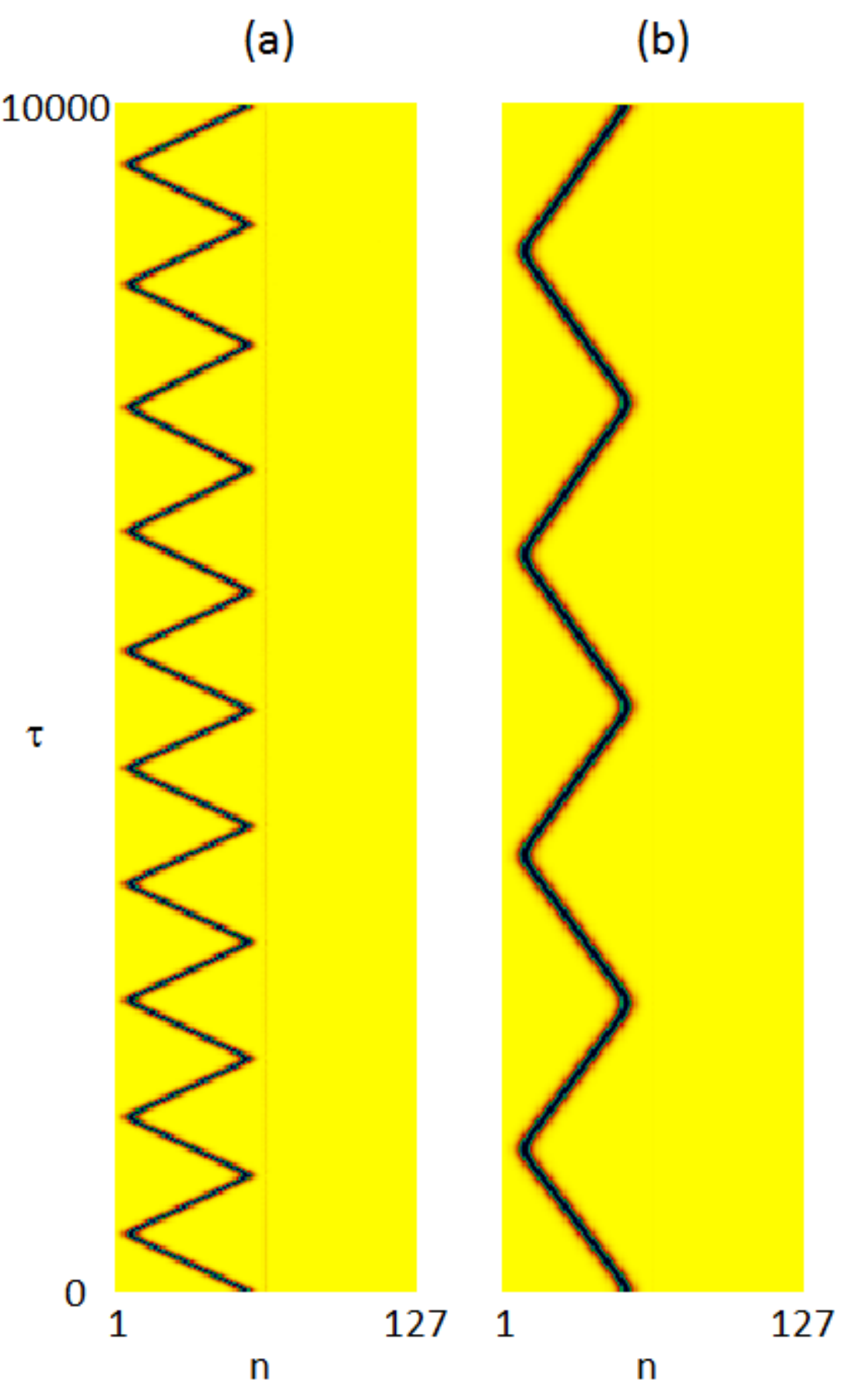}
\caption{\color{black}(Color online) Interaction between initially 
stationary breathers in the $^{168}$Yb + $^{170}$Yb mixture for (a)$D=8$ 
and (b)$D=11$. Both images show only the evolution of the $^{168}$Yb 
part of the mixture since the $^{170}$Yb evolution is rather straightforward, 
with the breather highly localised in the centre.  
The majority of $^{168}$Yb atoms forms a travelling breather while the remaining 
atomic density is absorbed by the $^{170}$Yb breather to form a symbiotic breather. 
The travelling breather reflects off the nearest boundary and then off the 
$^{170}$Yb breather in the centre of the lattice.  
\label{figured8d11}}
\end{figure}

It has been demonstrated that initially Gaussian wavepackets can evolve via 
the single species DNLSE into static breathers \cite{trombettoni01,franzosi11,neff14}. 
If the wavepacket is given a momentum in a certain direction, travelling 
breathers that translate across the lattice can also be formed. The general 
expression of the initial wavepacket is:
\begin{equation}
  z_{i,n} = \sqrt{\frac{1}{\sqrt{2\pi \sigma_i^2}}}
  \exp{\left(-\frac{(n-\bar n_i)^2}{4\sigma_i^2}\right)} \;\; e^{ip_in}
  \label{gaussian}
\end{equation}
where $\sigma_i$ is the initial width of the Gaussian cloud, and $\hat n_i$ is its 
position. For the single spcies case, low nonlinearity and values of $|p_i|$ between 
zero and $\pi/2$, corresponding to a positive $\cos p_i$, the cloud expands diffusively 
within the lattice. Localization into static breathers is then observed when increasing 
the repulsive self-interaction $\Lambda_{i,i}$. However, when the pseudomomentum crosses 
$\pi/2$ and the repulsive self-interaction $\Lambda_{i,i}$ is not too large, travelling 
breathers are formed (unless $\cos p_i$ is exactly 1, in which case the breather is 
stationary) \cite{trombettoni01,gomez04,livi06,franzosi07,franzosi11,neff14}.

In this section, we run simulations starting from stationary breathers of 
$^{168}$Yb and $^{170}$Yb in separate positions. These are formed by running 
simulations of initially Gaussian wavepackets with $\Lambda_{1,2} = 0$ and letting 
them reshape with dissipation applied at the boundaries to get rid of the background 
(see \cite{franzosi11} for a detailed description of the effects of dissipation). 
Once the breathers are formed and the background noise has vanished, we turn dissipations 
off and set $\Lambda_{1,2}$ to the value of 2.51.  

For each simulation, we change the initial distance $D$ between the centres of the 
single-species breathers and then study the breather interaction. With the stationary 
breathers centred on the same site at the start of the simulation ($D=0$), a symbiotic 
breather is formed, with the $^{168}$Yb breather expelling excess atomic density into 
the background (see Fig.~\ref{figuredensprof} (a) and (b)). The density profile of the 
symbiotic breather is then different than that of the two breathers occupying the same 
site with $\Lambda_{1,2} = 0$.

This behaviour keeps occurring when the initial distance between the breathers 
is larger, but still small enough that the density profiles overlap significantly.  
An example of this behavior is shown in Fig.~\ref{figuredensprof} (c) and (d), 
for $D=2$. Here, we see that less of the $^{168}$Yb atoms join with the symbiotic 
breather and more are expelled into the background. A small travelling breather is 
then formed from the atoms in the background. The reshaping process of the $^{170}$Yb 
breather is much the same as with $D=0$.

The formation and evolution of a traveling breather out of the interaction of two static 
breathers of separate species is shown more clearly in Fig.~\ref{figured8d11}, in which 
we see that the travelling breather reflects off the nearest boundary and then off the 
breather in the center of the lattice. In all simulations we consider that lattice sites 
outside the condensate are empty resulting in reflective boundaries. This is realized 
experimentally by fixing the size of the condensate with external magnetic fields. 
This is why breathers are observed to "bounce" off the boundaries in many figures 
of the paper. With $D=8$, the example shown in Fig.~\ref{figured8d11}(a), 
the initial density profiles of the breathers only overlap at the tails. The result 
of this is that only a small density of $^{168}$Yb atoms contributes to the symbiotic 
breather. As $D$ is increased, more $^{168}$Yb atoms go into the travelling breather 
and less in the symbiotic one.  In Fig.~\ref{figured8d11}(b), for $D=11$, we see that 
the symbiotic breather is not formed anymore and the travelling breather is denser, 
and slower, than that formed for $D=8$ (Fig.~\ref{figured8d11}(a)). This is due to 
the initial density profile for $D=11$, in which the breathers do not overlap any 
longer. It is importatn to stress that the motion of the travelling breather made of 
$^{168}$Yb atoms is due to the interaction of the two species via $\Lambda_{1,2}$ being 
different from zero. In the case of no-interaction ($\Lambda_{1,2} = 0$) both breathers 
remain stationary at all times. For all these examples, the evolution of the $^{170}$Yb 
breather is very similar to the case of $D=0$, with the high localisation in the centre 
of the lattice and minimal background noise.



\section{Collision of travelling and stationary breathers}

\begin{figure}[t]
\includegraphics[angle=0.0,clip,width=0.32\columnwidth]{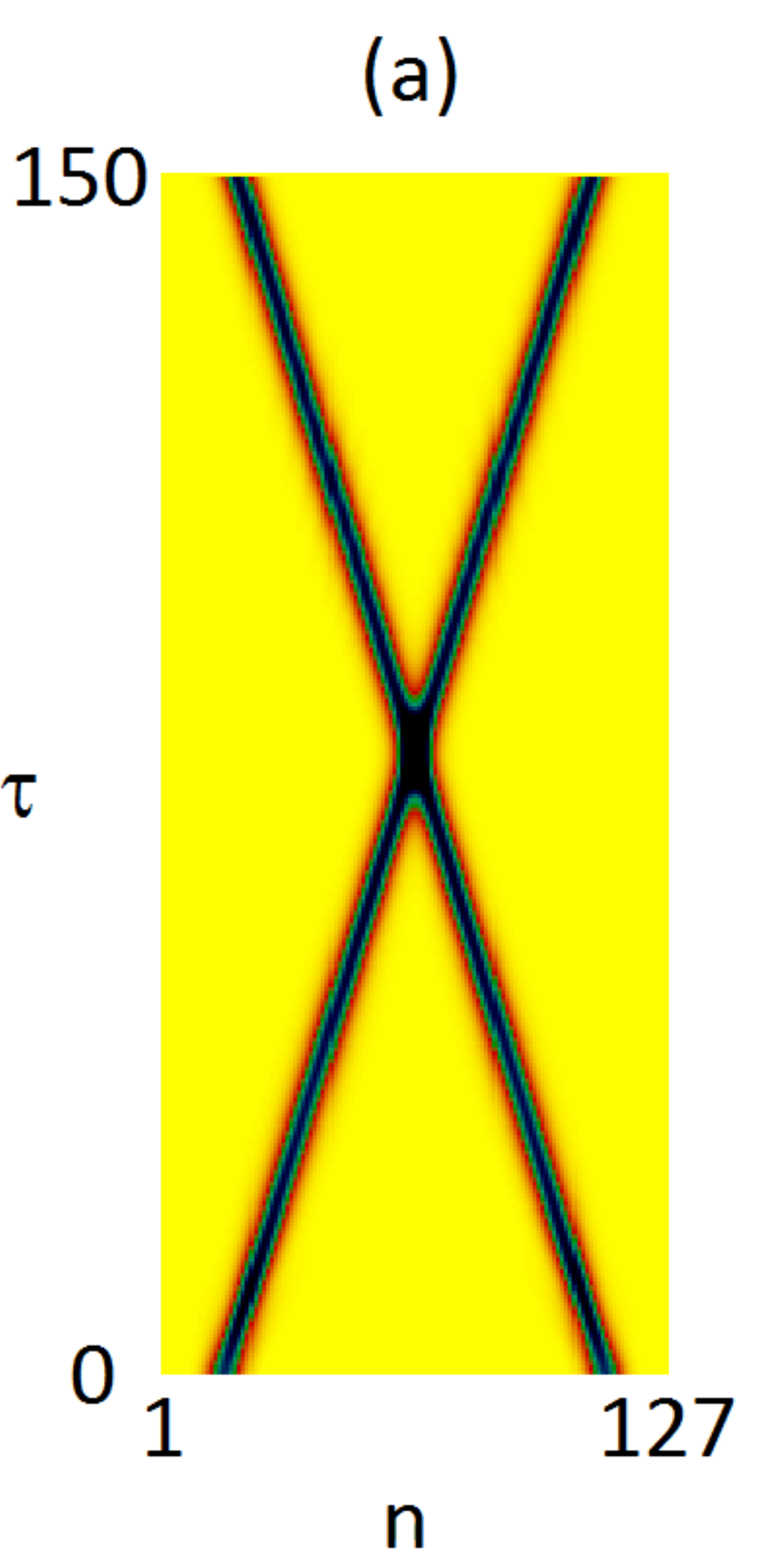}
\includegraphics[angle=0.0,clip,width=0.32\columnwidth]{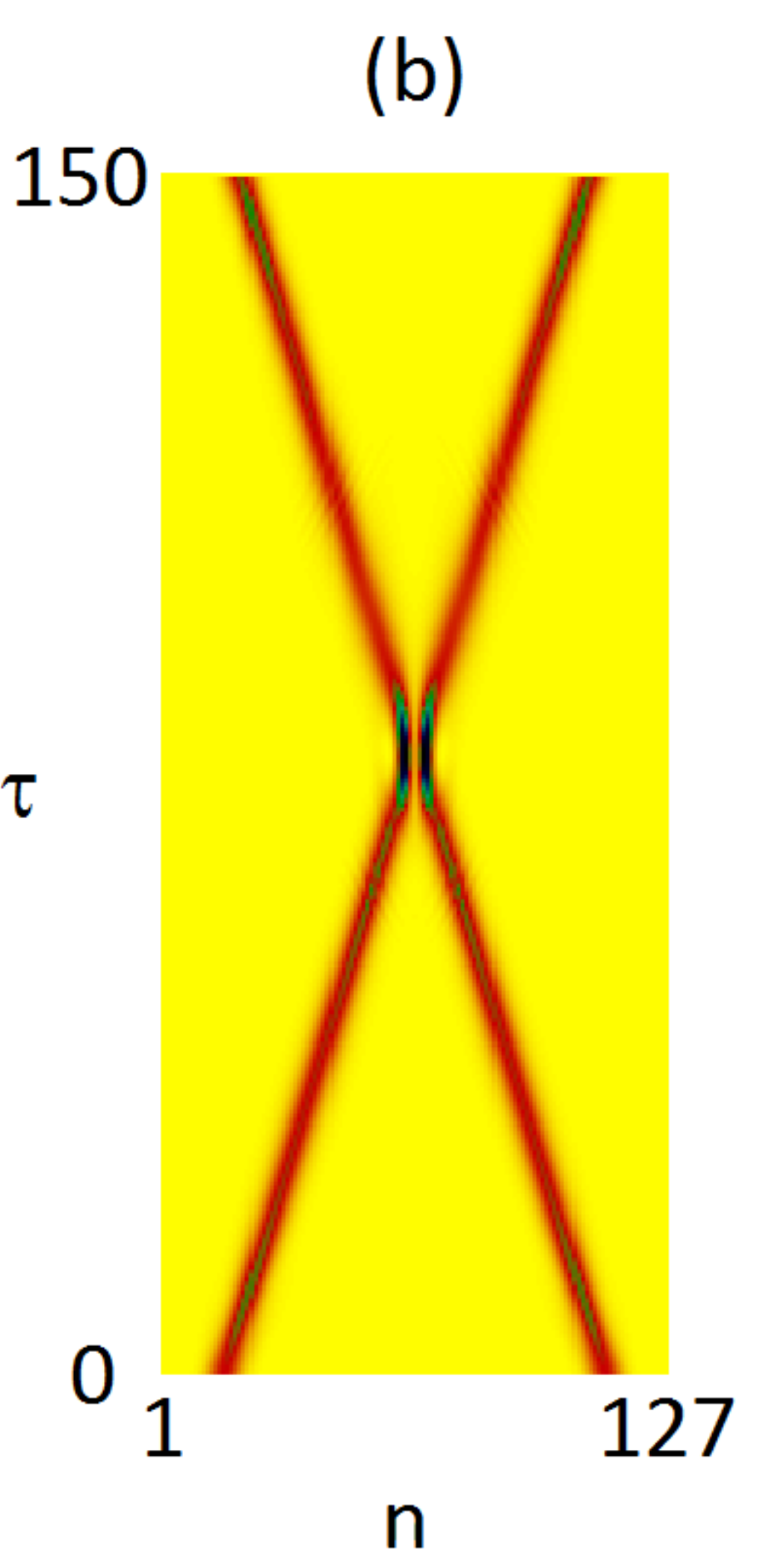}
\includegraphics[angle=0.0,clip,width=0.32\columnwidth]{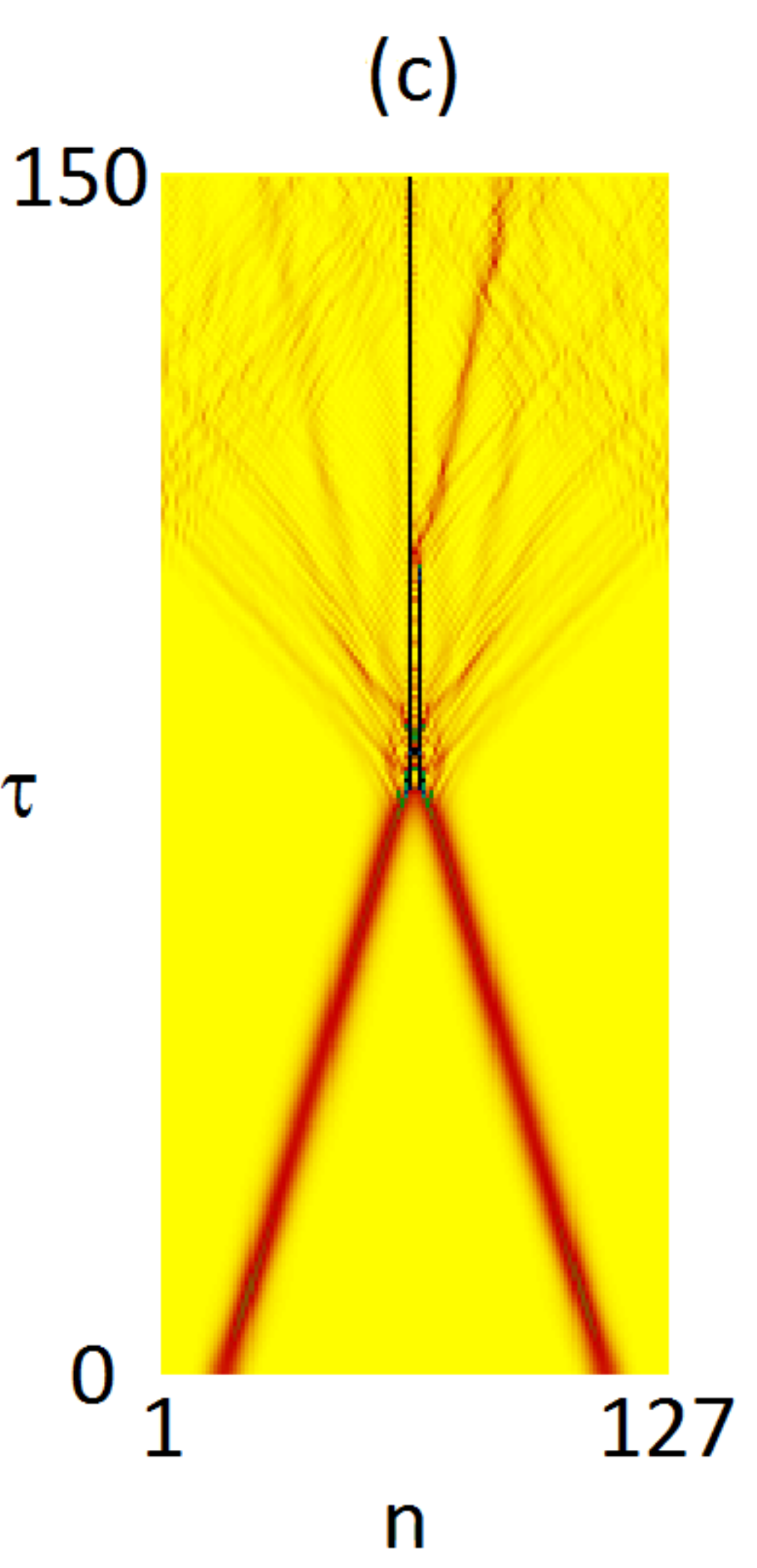}
\caption{\color{black}(Color online) Collision of two traveling breathers in the
  $^{170}$Yb (left) + $^{174}$Yb (right) mixture. The initial conditions are 
  $\Lambda_{1} = \Lambda_{2} = 1.1$, $\bar n_1 = 16$, $\bar n_2=112$, $\sigma_1 = \sigma_2 = 3$ 
  and $\cos p_1 = \cos p_2 = -0.95$ for all panels.  
  Note that $p_1 = - p_2$ and therefore the traveling breathers move in opposite directions. 
  $\Lambda_{1,2}=0$ is set to: (a)0, (b)-20 and (c)30. In (a), the breathers ignore each other 
  acting as if the other species was not present.  In (b), the breathers collide 
  elastically. In (c), the breathers are destroyed and a new symbiotic breather is created.
  \label{figure0}}
\end{figure}

\begin{figure}[t]
\includegraphics[angle=0.0,clip,width=0.9\columnwidth]{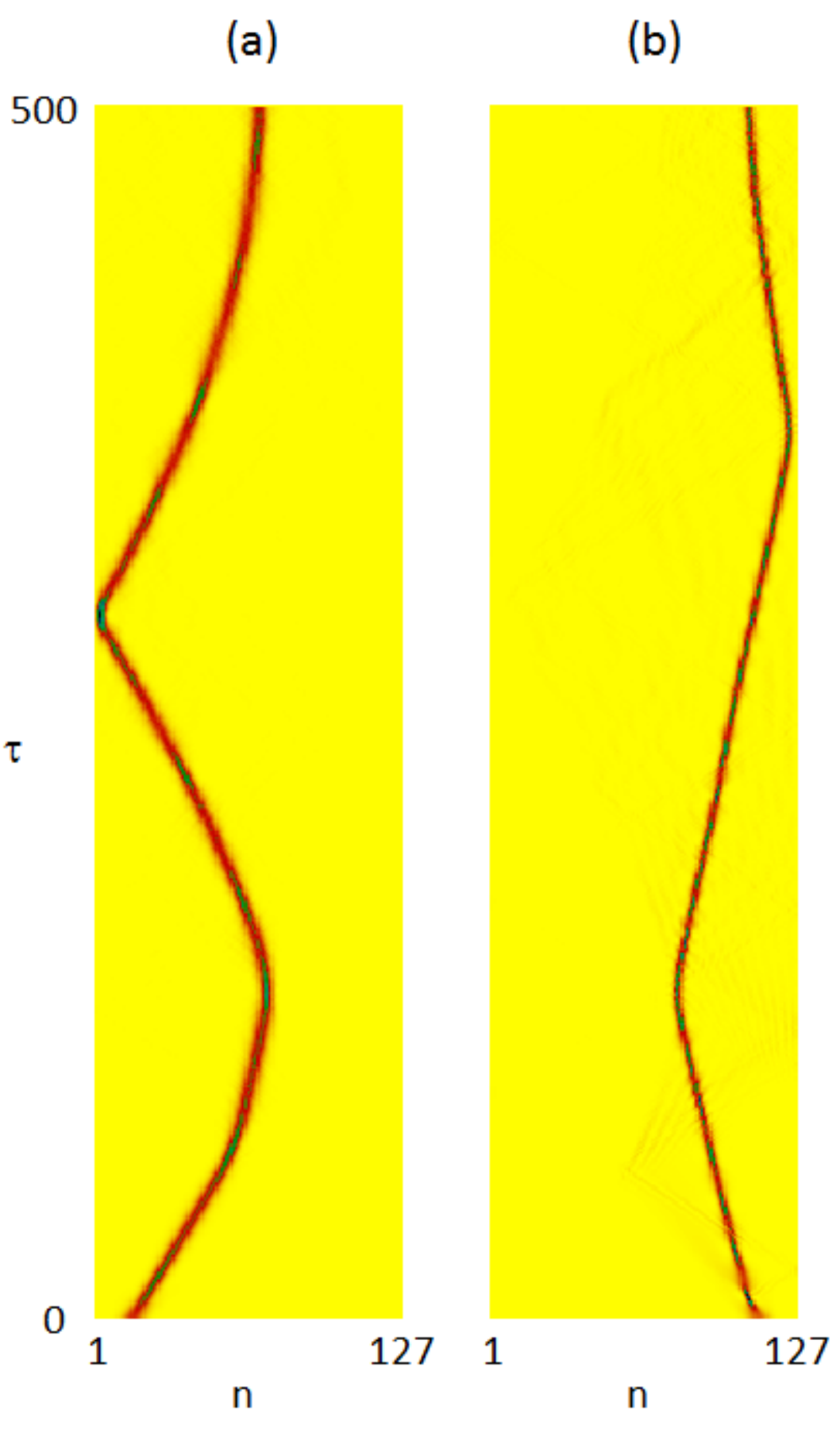}
\caption{\color{black}(Color online) A collision of two traveling breathers in the
  $^{170}$Yb (a) + $^{174}$Yb (b) mixture characterised by a large negative
  interspecies scattering length of -27.3~nm. The Gaussian parameters for the 
  initial condition are $\bar n_1 = 16$, $\bar n_2=112$, $\sigma_1 = \sigma_2 = 3$
  and $\cos p_1 = \cos p_2 = -0.95$.  The
  collision is elastic.\label{figure1}}
\end{figure}

\begin{figure}[t]
\includegraphics[angle=0.0,clip,width=0.9\columnwidth]{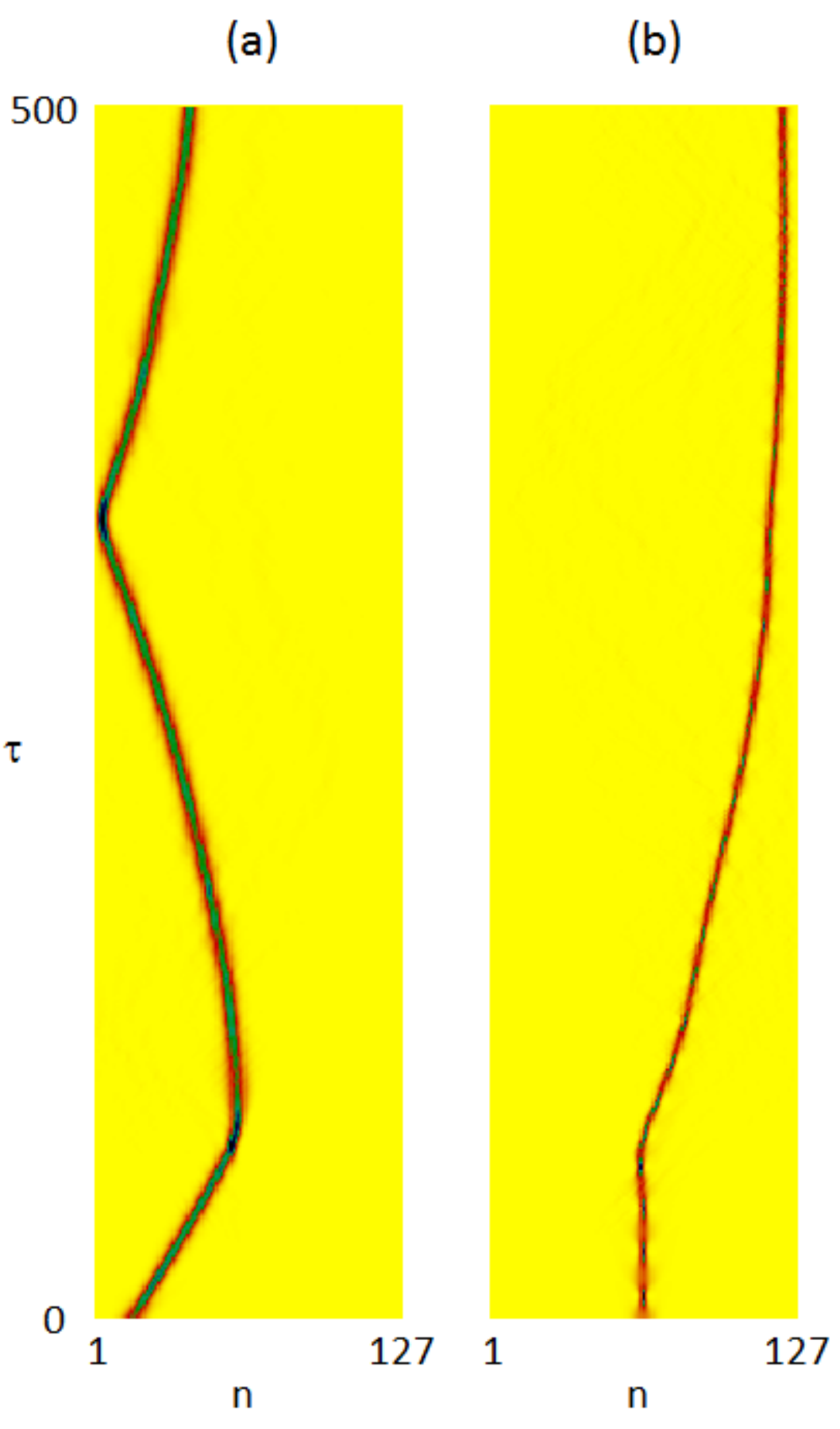}
\caption{\color{black}(Color online) A collision of a traveling
  breather (a) with a stationary breather (b). The physical situation is the
  same as in Fig. \ref{figure1}, except that $\bar n_2=64$ and
  $\cos p_2 = -1$ to make a stationary breather. The traveling breather 
  transfers large part of its (pseudo)momentum to the stationary one 
  and nearly stops. \label{figure2}}
\end{figure}

\begin{figure}[t]
\includegraphics[angle=0.0,clip,width=0.9\columnwidth]{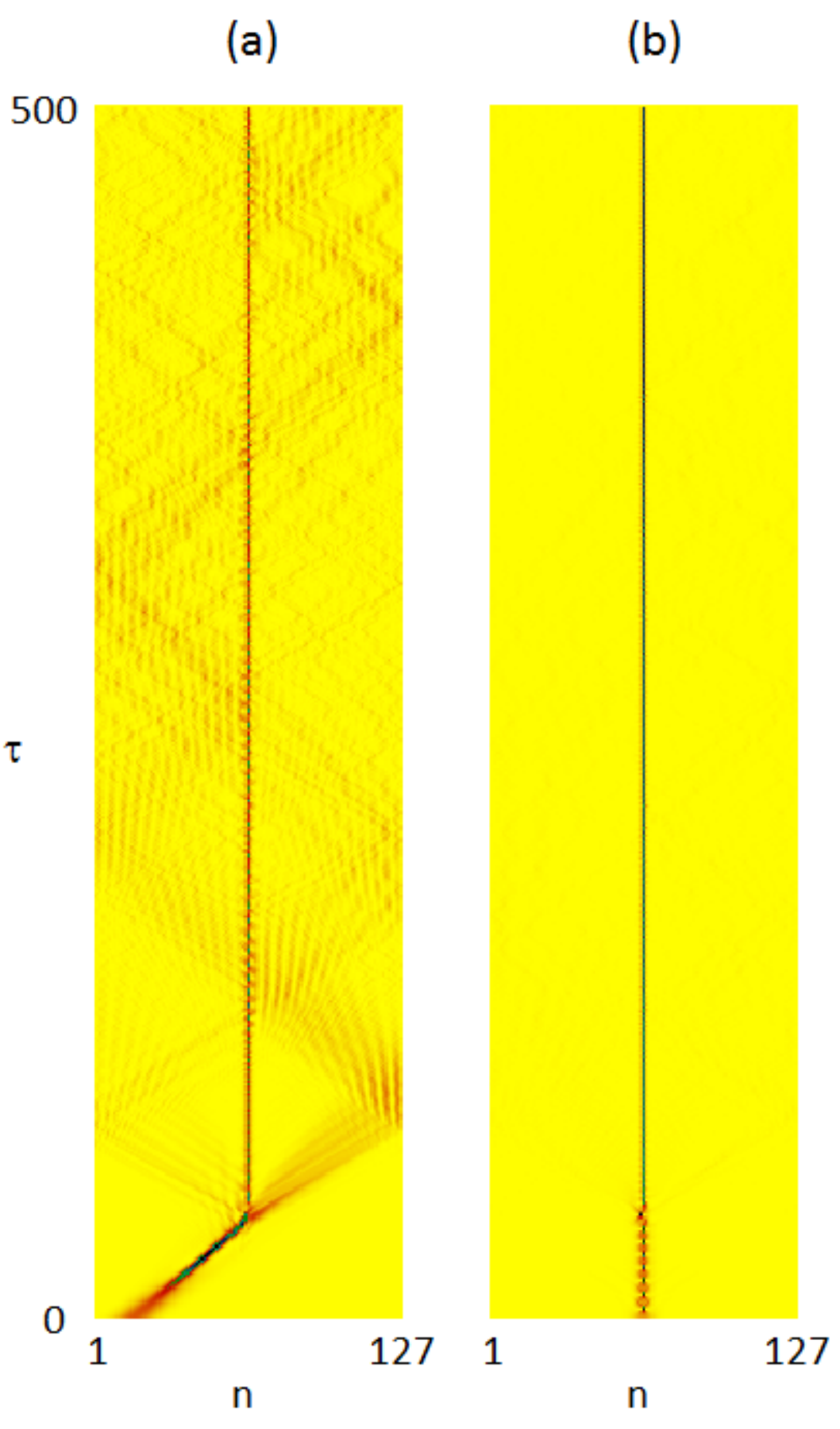}
\caption{\color{black}(Color online) A collision of a traveling breather 
  ($^{170}$Yb, shown in (a)) and a stationary breather ($^{168}$Yb, shown in (b)) for a
  positive interspecies scattering length of 6.2~nm. The initial condition
  parameters are $\bar n_1 = 16$, $\bar n_2 = 64$, $\sigma_1 = 5$,
  $\sigma_2 = 3$, $\cos p_1 = 0.8$, $\cos p_2 = 1.0$. As in Fig. \ref{figure0}(c), the traveling 
  breather is destroyed and a new \emph{symbiotic} soliton is
  created. \label{figure3}}
\end{figure}

\begin{figure}[t]
\includegraphics[angle=0.0,clip,width=0.9\columnwidth]{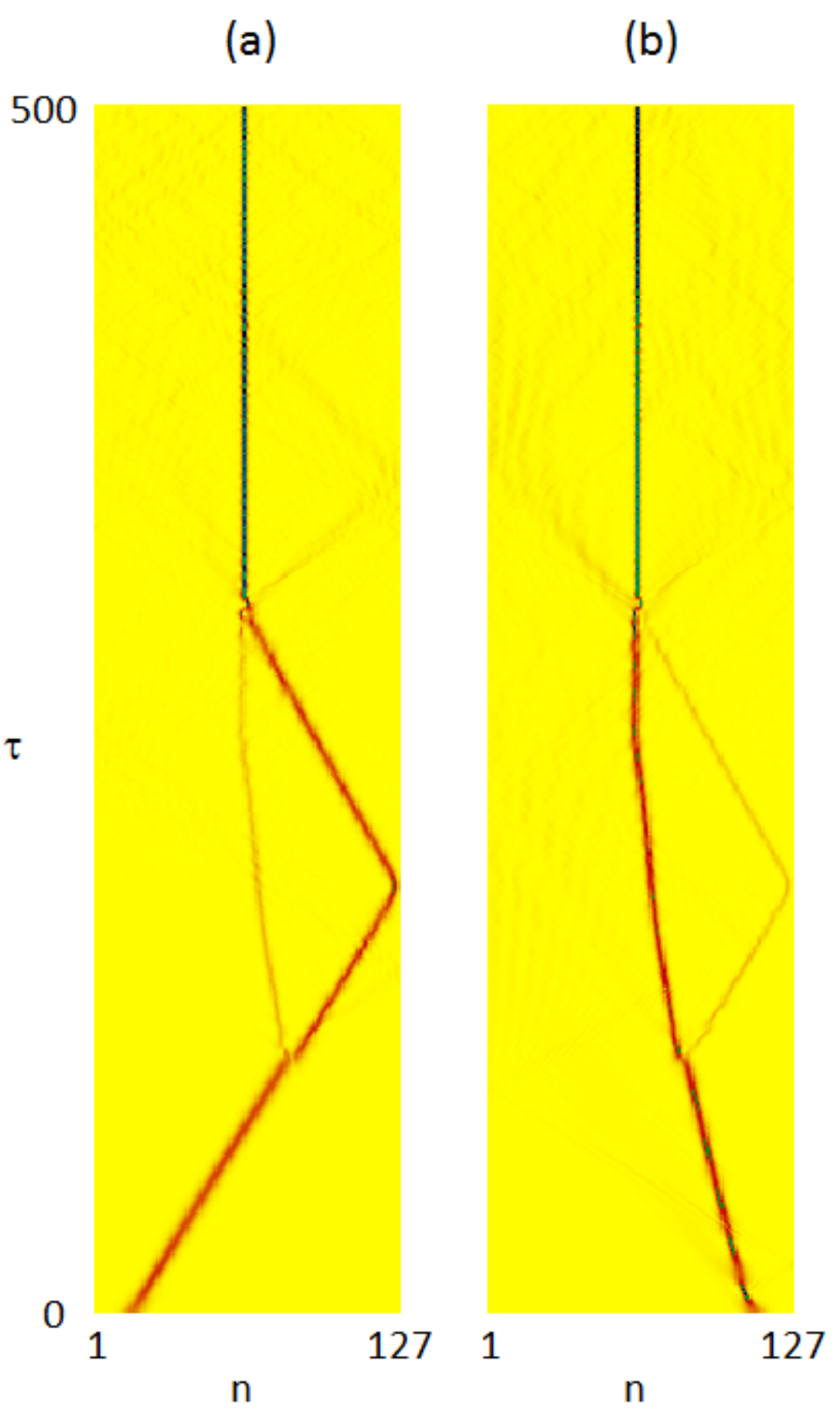}
\caption{(Color online) The same physical situation as in Fig.~\ref{figure1},
  except that the parameter $\Lambda_{1,2}$ describing the interspecies
  has now been increased to 4, corresponding to a positive
  interspecies scattering length of 8.9~nm. In this regime, the two
  breathers tunnel through each other. Note that a small part of each
  breather is trapped inside the other forming double-species
  traveling breathers.\label{figure4}}
\end{figure}

\begin{figure}[t]
\includegraphics[angle=0.0,clip,width=0.9\columnwidth]{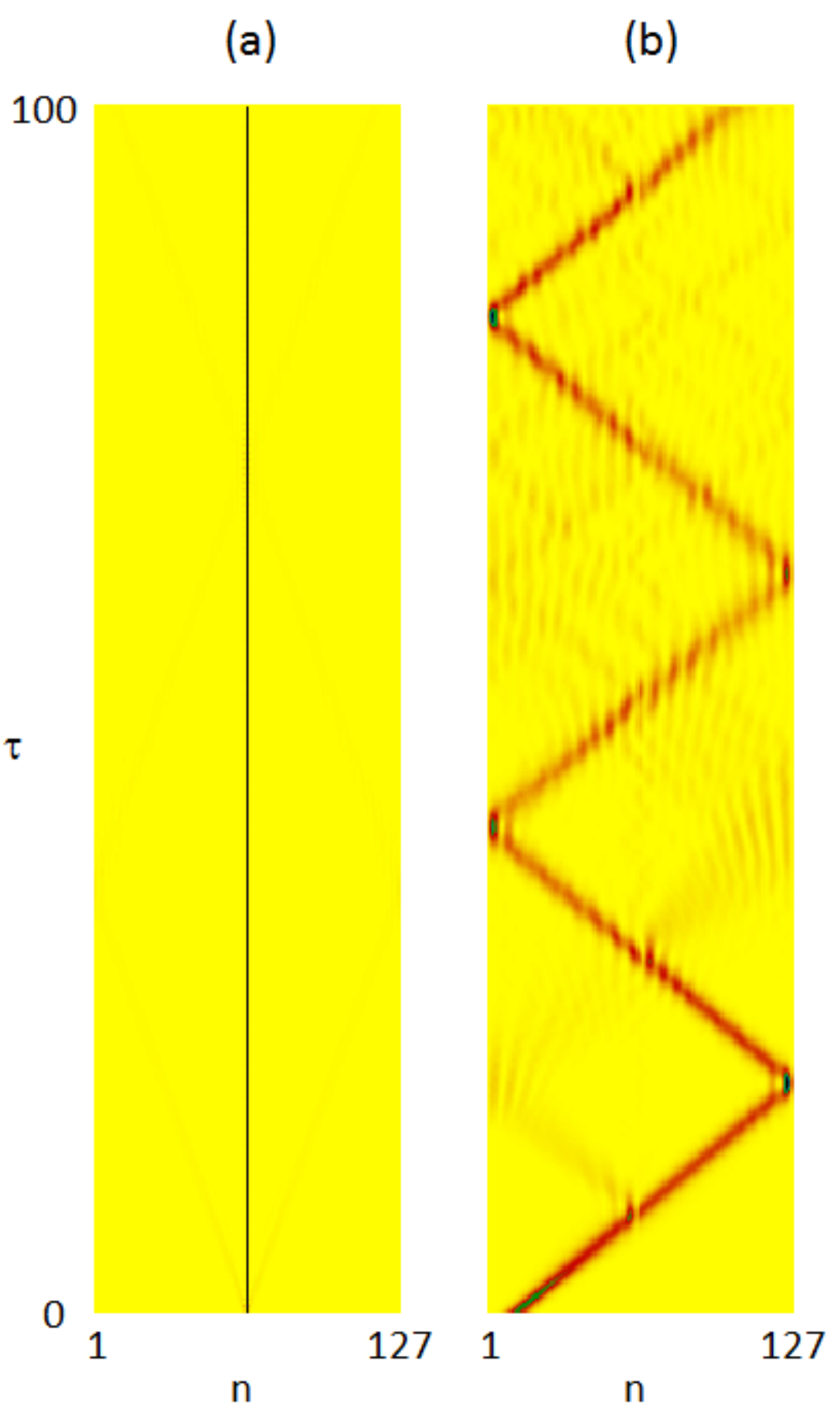}
\caption{(Color online) A collision of a traveling ($^{41}$K, shown in (b)) 
  and a stationary breather ($^{87}$Rb, shown in (a)), with
  the interspecies interaction parameter $\Lambda_{1,2} = 3$. The
  initial condition parameters are $\bar n_1 = 64$, $\bar n_2 = 12$,
  $\sigma_1 = 0.5$, $\sigma_2 = 3$, $\cos p_1 = 1$, $\cos p_2 =
  -0.9$. The traveling breather tunnels almost completely through 
  the self-trapped state.\label{figure5}}
\end{figure}
\begin{figure}[t]
\includegraphics[angle=0.0,clip,width=0.9\columnwidth]{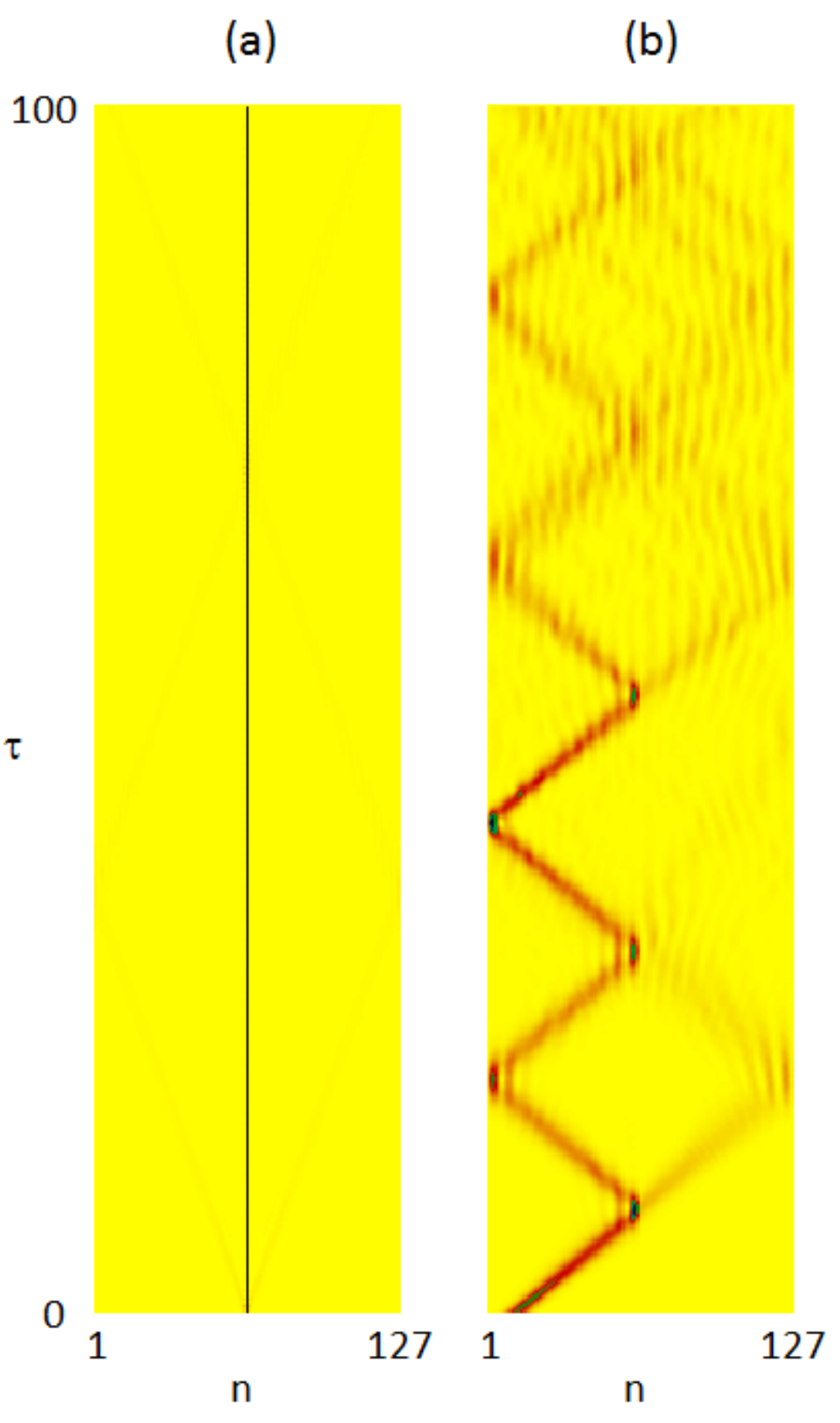}
\caption{(Color online) The same physical situation as in
  Fig. \ref{figure5} but with $\Lambda_{1,2} = -9$. The traveling
  breather bounces elastically from the self-trapped state with only
  a minor proportion tunneling through.\label{figure6}}
\end{figure}

Having assessed the interaction of discrete breathers of different atomic 
species when set in close proximity, we investigate here the collision of 
these breatehrs having set one or both of them in motion by using the 
pseudomomentum $p_i$ in the initial conditions given by (\ref{gaussian}). 
Experimentally, traveling breathers can be constructed by accelerating the 
lattice confining the condensate. Such accelerations were realised experimentally 
in \cite{fallani03} while observing optical lensing effects. 
In the case of two different species in the same lattice the difference in the 
masses will naturally lead to different pseudomomenta of the clouds after 
acceleration to the same group velocity, thus making collisions possible.
  
In Fig.~\ref{figure0}, we show three symplectic simulations of two colliding 
traveling breathers, with carefully chosen parameters so that there would be 
a minimal amount of sound-waves emitted from the initial Gaussian distributions.
In all simulations presented in Fig.~\ref{figure0} the breathers start from the 
same initial conditions but with different values of the interspecies interaction 
parameter $\Lambda_{1,2}$.  The color represents the (normalized) atomic density 
$|z_{n,i}|^2$ with $i=1$ for one species and $i=2$ for the other. The simulations 
start with two Gaussian wavepackets of the form (\ref{gaussian}) with 
$\cos p_1 = \cos p_2 = -0.95$ and $p_1 =-p_2$ to form two colliding traveling 
breathers.    

In Fig.~\ref{figure0}(a) when the interspecies interaction parameter $\Lambda_{1,2}$ 
is set to zero, the breathers follow the dynamics of single-species condensates. 
They are unaffected by the others presence and pass through each other.  When 
$\Lambda_{1,2}$ is a non-zero value, the two species affect each other when occupying 
the same lattice sites.  This is shown in Fig.~\ref{figure0}(b) and (c). At the beginning 
of the simulations, when the breathers are far apart from one another in the lattice, 
they follow the same path as in Fig.~\ref{figure0}(a), until they collide. For large 
negative values of $\Lambda_{1,2}$, the breathers collide elastically, as shown in figure 
\ref{figure0}(b). In this example, $\Lambda_{1,2}=-20$ and the breathers become narrower 
when they collide.  In Fig.~\ref{figure0}(c), $\Lambda_{1,2}$ is changed to a positive value 
and the collision is not elastic.  At the collision, the breathers explode, emitting a 
large amount of sound waves and a stationary symbiotic soliton composed of both species 
is formed.    

It is worth noting that elastic collision occurs when the interspecies interaction parameter 
is negative, which would normally imply \emph{attractive} interactions between the two species. 
Normally the two clouds try to achieve maximal overlap in order to minimize energy, while in 
our case the tendency is to minimize the overlap and retain separation of at least a few 
lattice sites. An explanation of this phenomenon by using the negative effective mass
of the discrete breathers is provided in Section V.


For the simulations in Fig.~\ref{figure0}, we have carefully chosen the parameters so that 
the amount of sound-waves emmited from the breathers is minimal.In the following we turn our 
attention to values of the parameters chosen from table \ref{parametertable} to model mixtures 
of ytterbium isotopes and that of $^{41}$K+$^{87}$Rb in realistic configurations. In these 
simulations, although the initial condition emits large amount of noise, we show that the 
main dependance of the collision from the interaction parameter $\Lambda_{1,2}$ remains that 
displayed in Fig.~\ref{figure0}.

In Fig.~\ref{figure1} we show a numerical simulation of two colliding traveling breathers in 
the $^{170}$Yb + $^{174}$Yb mixture that displays similar results to that of Fig.~\ref{figure0}. 
Note that for all following simulations, the dynamics of each species is shown in separate panels, 
unlike Fig.~\ref{figure0}. For example Fig.~\ref{figure1}(a) shows the $^{170}$Yb species, while 
Fig.~\ref{figure1}(b) shows the $^{174}$Yb species. At about $\tau \approx 90$ the two breathers 
collide elastically, as in Fig.~\ref{figure0}(b) since the interspecies scattering length is large 
and negative.

Figure \ref{figure2} shows a similar situation, except that now one of the breathers ($^{174}$Yb) 
is at first stationary ($\cos p_2 = -1$). After the collision the initially traveling breather 
(almost) stops while the other, up to now stationary, starts traveling. One could argue that this 
is a manifestation of a form of conservation of momentum. Again, as in Fig.~\ref{figure0}, the 
elastic behaviour occurs even though the $^{170}$Yb and $^{174}$Yb pair is described by a large 
negative scattering length of $a_{1,2} = -27.3$~nm, which, under normal circumstances,
stands for \emph{attraction} between the atoms of the two species. 

Compared to the examples in Fig.~\ref{figure0}, there is a much larger amount of sound waves 
emitted from the breathers in Figs.~\ref{figure1} and \ref{figure2} due to the parameter values 
used from Table \ref{parametertable}.  These small-amplitude sound waves do not appear to affect 
the main collision in a significant way and, in fact, a careful examination shows that the location 
of a breather of a certain species acts as an effective barrier to the sound waves of the other 
species emitted during the formation of the breathers. Sound waves that would normally expand 
over the entire lattice are now confined to a region limited by the position of the other-species 
breather.

A qualitatively different behavior from the above and similar to that in Fig.~\ref{figure0}(c), is 
found in the case of the $^{168}$Yb+$^{170}$Yb pair described by a positive (\emph{repulsive}) 
scattering length of $a_{1,2} = +6.2$~nm. Due to the large intraspecies interaction of $^{168}$Yb 
it is difficult to construct a clear traveling breather and therefore we limit ourselves to the 
case where the $^{168}$Yb breather is initially stationary. This situation is presented in 
Fig.~\ref{figure3}. As in Fig.~\ref{figure0}, at the time of impact ($\tau \approx 40$) the two 
breathers literally explode emitting a large amounts of sound waves and forming a double-species 
symbiotic stationary breather. In this state, the two component wavefunctions are well overlapped. 
Moreover, the final breather is much narrower than any of the original breathers before the collision. 
The frequency of oscillation of the two final co-located and co-existing breathers is species dependent. 
In the case displayed in Fig.~\ref{figure3} the frequency of the $^{168}$Yb breather is about 1.5 times 
that of the $^{170}$Yb breather located on the same site. 

To further explore these phenomena we carried out a simulation for the physical situation described in 
Fig.~\ref{figure1} of the $^{170}$Yb+$^{174}$Yb mixture but with the interaction parameter
$\Lambda_{1,2}$ increased to 4 corresponding to an interspecies scattering length of $a_{1,2} = 8.9$~nm. 
Such an increase can potentially be achieved using an optical Feshbach resonance. In this situation, we 
observe the collision of two traveling breathers with a positive interspecies interaction as shown in 
Fig.~\ref{figure4}. We find yet another collision behavior with these parameters: the two breathers 
mainly tunnel through each other but at each collision a fraction of the atomic species in one soliton 
becomes trapped inside the other. Moreover, the breathers appear to accelerate or decelerate for a brief 
time during the strong interaction.

Quite different results are obtained for the $^{41}$K+$^{87}$Rb mixture, characterized by a large 
tunneling rate ratio of $\gamma_2/\gamma_1 \approx 6.97$. This, together with the large self-interaction 
of Rb, changes the physics dramatically.

In Figs.~\ref{figure5} and \ref{figure6} we show a collision of a traveling breather and a self-trapped 
state of rubidium atoms. Within the range of our simulation parameters that simulates possible experimental 
realizations, it has proved not possible to initiate a traveling breather state with the Rb condensate due 
to its large self-interaction. We set the interspecies interaction parameter $\Lambda_{1,2} = 3$ for
Fig.~\ref{figure5} and $\Lambda_{1,2} = -9$ for Fig.~\ref{figure6}. In both cases the rubidium breather 
acts only as a potential barrier, through which some of the incoming potassium soliton can either reflect 
or tunnel. This behavior, which contrasts with the phenomena seen in the simulation with ytterbium, can 
be attributed to the drastically different tunneling rates of potassium and rubidium. 


\section{The collision mechanism}

\begin{figure}[t]
\includegraphics[angle=0.0,clip,width=\columnwidth]{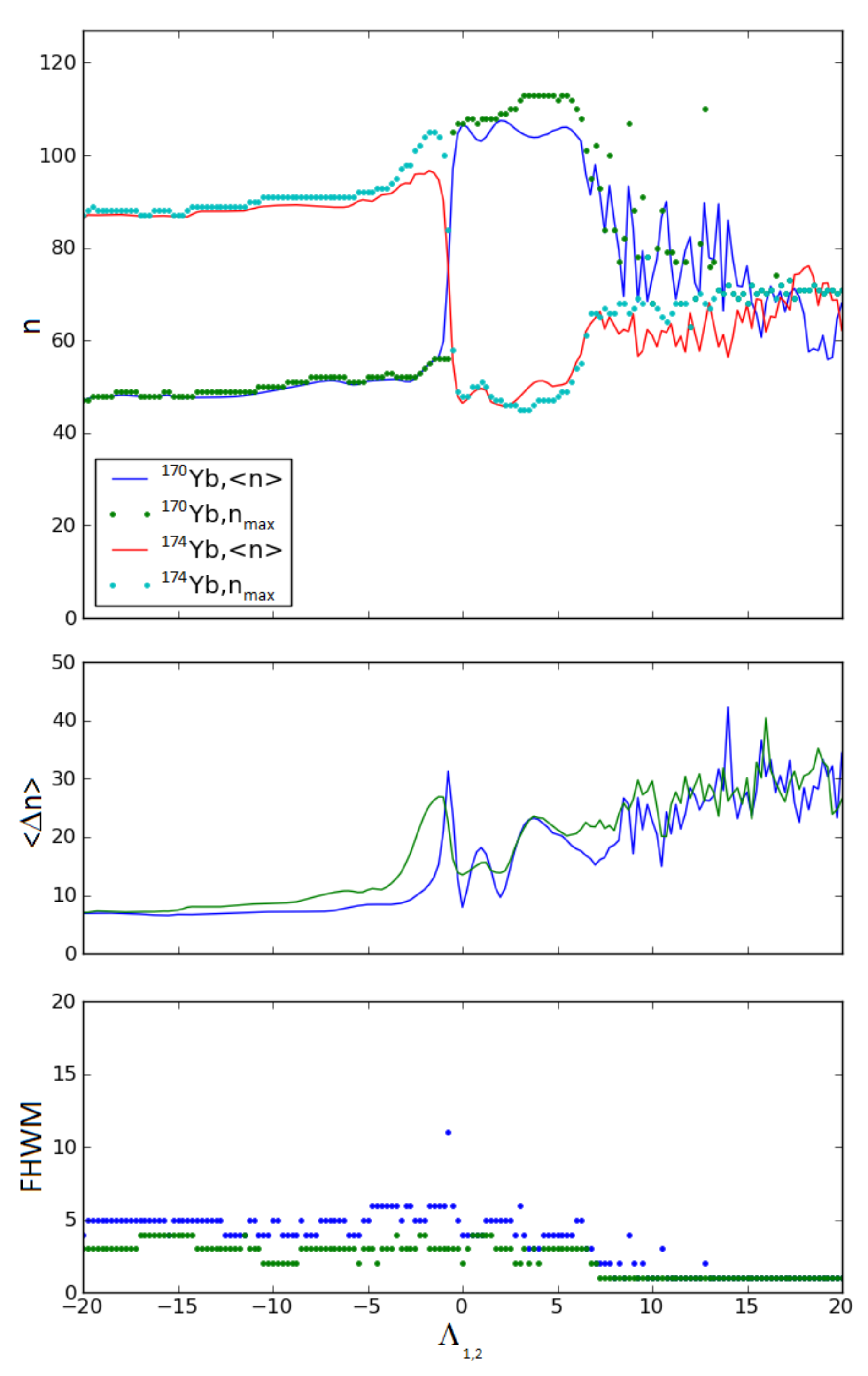} \caption{(Color
  online) The collision outcome as a function of the interspecies
  interaction, $\Lambda_{1,2}$. The top panel displays the mean and the peak 
  site per species, the center panel the standard deviation per species, and the 
  bottom panel the FWHM per species as defined in the text. Symplectic simulations 
  corresponding to the $^{170}$Yb+$^{174}$Yb mixture.
\label{figure8}}
\end{figure}

To understand the mechanisms at the base of the collision outcomes, we have investigated the 
dependence of the result of the collisional process with respect to the heteronuclear interaction.
By treating the discrete condensate wavefunctions as distributions we define the mean lattice site
\begin{equation}
    \langle n_i \rangle = \sum_{n}n\left|z_{i,n}\right|^2 \, ,
    \label{mean}
\end{equation}
and its standard deviation,
\begin{equation}
    \langle\Delta n_i\rangle = \left(\sum_{n}(n- \langle n_i \rangle)^2 
     \left|z_{i,n}\right|^2\right)^{1/2} \, .
    \label{stddev}
\end{equation}
These two parameters describe the \emph{global} behavior of the condensate. In fact, if a breather 
splits or is destroyed in a collision, the standard deviation increases dramatically. To assess
the \emph{local} behavior, i.e. looking for a new breather created in a collision, we also search 
for the site with the largest number of atoms, $n_{{\rm max},i}$ and attempt to estimate the new 
breather's width (Full Width Half Maximum, FWHM) by counting the adjacent sites which contain at 
least half the number of atoms of those in the site of the maximum.

Figure \ref{figure8} shows the parameters $\langle n_i \rangle$, $n_{{\rm max},i}$, $\langle \Delta n_i \rangle$ 
and the FWHM as a function of the mutual interaction parameter $\Lambda_{1,2}$. For each value of 
$\Lambda_{1,2}$ a simulation was performed up to $\tau = 150$, just past the collision. The initial 
conditions and interaction parameters are the same as in Fig.\ref{figure1}, except for the scanned 
$\Lambda_{1,2}$ and the tunneling ratio $\gamma_2/\gamma_1$, which is set to 1.

Four different regimes can be identified from Fig.~\ref{figure8}. On the left, for $\Lambda_{1,2}$ 
lower than about $-2.0$, two traveling breathers collide elastically and remain basically unaffected 
by the collision. This  is the situation shown in figure \ref{figure1}. Then, there is a transition 
point at $\Lambda_{1,2} \approx -1.5$ where the each breather splits into two as they collide. This 
results in a sudden increase of $\langle \Delta n \rangle$ for each species. Note that this increase 
differs between the two species; for $^{174}$Yb it's maximum is located at $\Lambda_{1,2} \approx -2$ 
as opposed to $\Lambda_{1,2} \approx -1$ for $^{170}$Yb, quite probably as a result of the different 
self-interaction parameters. 

For small, but positive values of $\Lambda_{1,2}$ (i.e. less than 6) the two breathers tunnel through 
each other. Note that $\langle \Delta n \rangle$ remains low in this regime (about 15 sites, growing 
slowly to 20) which means that the breathers are not destroyed. This is shown in Fig.~\ref{figure4} for 
$\Lambda_{1,2} = 4$. As the two breathers tunnel through each other, a part of their wavefunction is 
trapped inside the other soliton; this effect grows as the interspecies interaction increases leading 
to a slow increase in the standard deviation of the atomic density distributions.

A rather sudden change takes place at about $\Lambda_{1,2} = 6$. The  system becomes visibly sensitive 
to small changes in the mutual interaction. This is the regime where the collision results in the
destruction of the two breathers. Fig.~\ref{figure3} is an example of such a case. The process is chaotic, 
yet in many cases leads to the creation of a double-species symbiotic breather manifested by its very 
low FWHM.

  
\subsection{Discussion}
       
The presented results can be reasonably explained by using one of the key phenomena at the base of 
gap solitons: the \emph{negative effective mass}. The dynamics of a gap soliton in an external potential 
are exactly opposite to what one would expect - a gap soliton attempts to climb potential hills and 
in itself is a balance between its negative effective mass that tries to make it collapse and its 
repulsive self-interaction that prevents it \cite{trombettoni01,matuszewski07}. In fact, the variational 
model of a wavepacket used in Trombettoni \emph{et al.} \cite{trombettoni01} shows that the wavepacket 
center obeys a Newton-like dynamics when $p \approx 0$ and exactly the contrary when $p \approx \pi$.

This `contrary' behavior of the solitons seems to be the key to the explanation of our findings. 
The totally elastic collision encountered when the interspecies interaction is highly \emph{attractive} 
would be caused by the fact that the solitons `see' each other as potential walls rather than wells.

The splitting behavior has been investigated in a slightly different context by Matuszewski \emph{et al.} 
\cite{matuszewski07} where the dynamics of two already overlapped stationary solitons was analyzed. In our 
case splitting happens if the attractive interaction is small enough to let the two breathers overlap 
briefly. Then the system becomes unstable and each breather splits in two. It is also possible to look 
at this phenomenon from a different angle. Due to its negative effective mass, the split of the breather 
is quite similar to the case of a wavepacket encountering a potential barrier where, depending on the 
barrier height (or the interaction between the breathers) part of the wavepacket goes through while the 
rest is reflected.

In the repulsive interaction regime the breathers behave as if they saw each other as potential wells. 
This is again an effect of their negative effective mass and, consequently, reversed dynamics. Thus,
for the collision's duration, their speed increases (at the cost of wavepacket spreading and of reducing 
their energy due to atomic self-interaction).

The chaotic behavior when the interspecies interaction is large and repulsive is probably caused by the 
system entering an unstable regime as predicted by Gubeskys \emph{et al.}\cite{gubeskys06}. The chaotic 
dynamics would then cause the destruction of the original two breathers and possibly the creation of a 
stable \emph{intragap} soliton. It is beyond the capabilities of our model to establish if it is possible 
for an \emph{intergap} soliton to emerge during the collision since the tight-binding approximation is 
limited to the lowest band-gap by definition.



\section{Conclusion}

We have analyzed the behavior of interacting and colliding discrete breathers in BEC composed 
of different atomic species in optical lattices. We have found that the interaction depends 
on the initial distance of the two breathers and led either to the formation of a symbiotic 
solitons or to the setup in motion of one of the two breathers. The collision outcome depends 
both on the tunneling rate ratios of the two species, as well as the interspecies interactions. 
When the tunneling rates differ greatly, as in the $^{41}$K+$^{87}$Rb mixture, one of the
breathers acts as an effective potential wall to the other and the whole process can be viewed 
as a case of one-particle scattering on a potential wall. 

In the case where the tunneling rates are comparable (like in the case of mixtures of Ytterbium 
isotopes) we have identified four collision regimes.  For large negative scattering lengths the 
collision is elastic and the two traveling breathers remain intact, with considerable momentum 
transfer between the two. For small negative scattering rates, the breathers overlap briefly and 
split in two, as originally predicted in \cite{matuszewski07}. When the interspecies interaction 
is weakly repulsive, the two breathers tunnel through each other unharmed for a wide range of 
interspecies interacions. Finally, with the interspecies interaction sufficiently large, the 
dynamics becomes chaotic and the two breathers are destroyed with a possible creation a new 
two-component soliton similar to an \emph{intragap} soliton as predicted in \cite{gubeskys06}. 
Feasible explanations to the above phenomena have been provided using the concept of negative 
effective mass and the resulting reversed dynamics.

Interaction and collision properties of localized excitations in BEC in optical lattices can 
have interesting applications in the realization of ultracold Bose-Fermi mixtures where gap 
solitons can be viewed as matter-wave counterparts of quantum dots and antidots \cite{salerno05}. 
Changing the species interaction allows one to tune the character of the collisions from fully 
elastic to fully inelastic and/or tunneling with clear advantages in the manipulation of 
information in matter-wave systems.

\section*{Acknowledgements}
We are indebted to Stefano Iubini, Massimo Inguscio, Francesco Minardi, Roberto Franzosi 
and Yoshiro Takahashi for useful discussions. This work was partially supported 
by the Institute of Complex Systems at Strathclyde and an EPSRC doctoral training 
fellowship. The research is part of the program of the National Laboratory FAMO in 
Toru\'n, Poland and partially supported by the Polish MNISW (Project No. 
N N202 1489 33).

\end{document}